\documentclass[12pt]{article}
\newcommand\be{\begin{equation}}
\newcommand\ee{\end{equation}}
\newcommand\ba{\begin{eqnarray}}
\newcommand\ea{\end{eqnarray}}
\newcommand\eq{\begin{equation}}           
\newcommand\en{\end{equation}}

\usepackage{ulem} % \sout{this is wrong} to draw the line on top of the sentence

\input epsf
\usepackage{amssymb}
\usepackage{amsmath}
\usepackage[dvips]{graphicx}
\usepackage{bm}
\usepackage{graphicx}
\usepackage{color}
\usepackage{subfigure}

\begin{document}
\title{
%{\hfill  \small\\ ~\\~\\}
The Effects of Dark Matter-Baryon Scattering on Redshifted 21 cm Signals}
 \author{Hiroyuki Tashiro$^{1}$, Kenji Kadota$^{2}$ and Joseph Silk$^{3,4,5}$\\
{ \small $^1$ \it   Department of Physics, Nagoya University, Nagoya
 464-8602, Japan} \\
% { \small $^2$ \it   Program for Leading Graduate Schools ``PhD
% professional:Gateway to Success in Frontier Asia'',}\\
% {\small \it Nagoya University, Nagoya 464-8602, Japan} \\
{ \small $^2$ \it  Center for Theoretical Physics of the Universe, Institute for Basic Science, Daejeon 305-811, Korea} \\
{ \small $^3$ \it Institut d'Astrophysique de Paris, CNRS, UPMC Univ Paris 06, } \\ 
{ \small \it UMR7095, 98 bis, boulevard Arago, F-75014, Paris, France} \\
 { \small $^4$  \it The Johns Hopkins University, Department of Physics and Astronomy, Baltimore, Maryland 21218, USA}\\
 { \small $^5$  \it Beecroft Institute of Particle Astrophysics and Cosmology, University of Oxford, Oxford OX1 3RH, UK}
}
\date{\vspace{-5ex}}
%\date{}  % Toggle commenting to test
\maketitle   

\begin{abstract}
We demonstrate that elastic scattering between dark matter (DM) and
baryons can affect the thermal evolution of the intergalactic medium at
early epochs and discuss the observational consequences. We show that,
due to the interaction between DM and baryons, the baryon temperature is cooled after decoupling from the CMB temperature.
We illustrate our findings by calculating the 21 cm power spectrum in
coexistence with a velocity-dependent DM elastic scattering cross
section. For instance, for a DM mass of $10$~GeV, the 21 cm
brightnesstemperature angular power spectrum can be suppressed by a
factor 2 within the currently allowed DM-baryon cross section bounded
by the CMB and large-scale structure data. This scale-independent
suppression of the angular power spectrum can be even larger for a
smaller DM mass with a common cross section (for instance, as large as
a factor 10 for $m_d\sim 1$ GeV), and such an effect would be of great interest for probing the nature of DM in view of forthcoming cosmological surveys.       
\end{abstract}

%\pacs{95.35.+d}
% 11.30.Pb Supersymmetry
% 12.60.Jv Supersymmetric models
% 95.30.Cq Elementary particle processes
% 95.35.+d Dark matter
% 98.35.Gi Galactic halo (Milky Way)
% 98.35.Df Kinematics, dynamics, and rotation
% 98.35.Pr Solar neighborhood
% 96.40.-z Cosmic rays (in the solar system)
% 98.70.Sa Cosmic rays (outside the solar system)
\maketitle   

\setcounter{footnote}{0} 
\setcounter{page}{1}
\setcounter{section}{0} \setcounter{subsection}{0}
\setcounter{subsubsection}{0}
\section{Introduction}

The nature of dark matter~(DM) is one of the greatest mysteries of modern
cosmology. One can infer its properties through its interactions with
other visible objects. Even though  conventional DM models assume only
gravitational interactions with ordinary baryonic matter, other forms of
couplings are not ruled out and  deserve further study in view of the
potential signals observable in forthcoming experiments.
DM-baryon interactions are of great interest for cosmology because the DM-baryon coupling can modify the evolution of structure
formation at early epochs, and  stringent constraints have been obtained
from  current data (e.g. CMB and Ly-$\alpha$) for a wide variety of dark
matter models such as  millicharged DM, dipole DM and strongly
interacting DM \cite{sper, chen, kam,mel, zurekyu, mark1}.

In this paper, we focus on the impact of the DM-baryon coupling on the
temperature evolution of DM and baryons and explore the consequences for
the redshifted 21~cm signal from very early epochs.
In the standard cosmology, the
baryon temperature $T_b$ couples with the CMB temperature $T_{\gamma}$ due to Compton
scattering via the small residual fraction of free electrons left over from  recombination 
down to a redshift $z_{\rm dec}(\sim 200)$ while $T_b$ subsequently
cools adiabatically at  lower redshift $z \lesssim z_{\rm dec}$. 
On the other hand, the DM temperature $T_d$ decouples from  $T_{\gamma}$ at a much earlier stage of the universe and $T_d$ is assumed to 
evolve adiabatically since then. The DM is hence ``cold'', and $T_d$ is much lower than $T_b$. Due to  DM-baryon coupling, however, the baryons can be cooled by the DM after the baryon temperature decouples from the CMB temperature. In order to probe this effect, we consider the
observations of redshifted 21~cm lines from neutral hydrogen during the dark ages before  reionization starts ($20 \lesssim z \lesssim 1000$).
The signal of redshifted 21~cm lines depends on the properties of baryon
gas at high redshifts: including  the
density, the temperature and the ionization fraction~\cite{Madau:1996cs}~(see
refs.~\cite{fur, Pritchard:2011xb} for recent reviews). The observations of redshifted~21 cm lines hence can provide  a probe of
the thermal evolution of baryonic gas. There have been related paper s investigating the
21~cm signal due to energy injection during the dark ages including the
dissipation of magnetic fields~\cite{Tashiro:2006uv,Shiraishi:2014fka},
energy injection from primordial black holes~\cite{Mack:2008nv,Tashiro:2012qe}, and the decay or annihilation of dark matter~\cite{Furlanetto:2006wp,Chuzhoy:2007fg}. Our study in contrast looks into the effects of elastic scattering between the DM and baryons on the 21 cm signals by quantifying the change in the evolution of $T_{b}$ and $T_{d}$ due to  DM-baryon coupling.

There are several on-going and planned projects to
measure the redshifted 21~cm signals by large interferometers such as
the LOw Frequency ARray~(LOFAR)~\cite{vanHaarlem:2013dsa}, the Murchison Widefield Array~(MWA)~\cite{Lonsdale:2009cb},
the Giant Metre-wave Radio Telescope~(GMRT)~\cite{Paciga:2013fj} and Square Kilometer Array~(SKA)\footnote{http://www.skatelescope.org/}. 
The purpose of this paper is to demonstrate the potential
significance of  DM-baryon coupling on the 21cm observables 
and investigate the range of DM-baryon coupling for  observational feasibility.

 We discuss, for simplicity, the case where cold dark matter accounts for the entire  DM density, and we calculate the 21 cm signal in the presence of DM-baryon coupling during the dark ages before  reionization starts $20 \lesssim z \lesssim 1000$ for its observational feasibility. This suffices for our purpose of quantifying the significance of DM-baryon coupling on future cosmological observables. 
Throughout this paper, we adopt the standard $\Lambda$CDM model parameters: $h=0.7$, $h^2 \Omega_b =0.0226$ and $\Omega_d=0.112$, where
$h$ is the present Hubble constant normalized by 100~km/s/Mpc and
$\Omega_b$ and $\Omega_d$ are the density parameters of baryons and DM.

\section{Thermal evolution of baryons and DM with DM-baryon coupling}

We solve the Boltzmann equations to follow the background temperature evolution.
The coupling between baryons and DM induces momentum transfer between
them, and the temperatures of DM and baryons, $T_d$ and $T_b$, evolve as~\cite{cm1} 
\ba
(1+z) \frac{d T_d}{dz} &=& {2} T_d + \frac{2m_d}{m_d+m_H}
\frac{K_b}{H}(T_d-T_b), 
\label{eq:Td_z} \\
(1+z) \frac{d T_b}{dz} &=& {2} T_b +
 \frac{2\mu_b}{m_e} \frac{K_{\gamma} }{H}(T_b-T_{\gamma})
 + \frac{2\mu_b}{m_d+m_H}\frac{\rho_d}{\rho_b} \frac{K_b} {H}(T_b-T_d),
\label{eq:Tb_z}
\ea
where
$\mu_b \simeq m_H (n_{\rm H}+4n_{\rm He})/(n_{\rm H}+n_{\rm He}+n_e)$ is the mean molecular
weight of baryons (including free electrons, and H, He ions), and 
$K_\gamma$ and $K_b$ are the momentum transfer rates. $K_\gamma$ represents the usual Compton collision rate
\ba
 K_{\gamma}= \frac{4 \rho _\gamma }{3 \rho_b} n_e \sigma_T,
 \ea
 where $\sigma_T$ is the Thomson scattering cross-section. For $K_b$, we
 consider the general form of cross section which can be velocity
 dependent parameterized by the baryon-DM relative velocity $v$ 
 \ba
 %\bar{\sigma}(v)=\sigma_0 v_{\rm rel} ^n.
 \sigma(v)=\sigma_0 v ^n,
 \ea
 so that the momentum transfer rate $K_b$ becomes \cite{mark1}
 \ba 
 K_{b}=\frac{c_n \rho_b \sigma_0}{m_H+m_d} \left( \frac{T_b}{m_H}+\frac{T_d}{m_d}\right)^{\frac{n+1}{2}}.
 \label{eq:thermal_coupling}
 \ea
The spectral index $n$ depends on the nature of DM models, for instance, 
$n=-1$ corresponds to the
Yukawa-type potential DM~, $n=-2,-4$ are respectively for dipole DM and
 millicharged DM~~\cite{kam,mel,zurekyu,mark1,davidson,feld,mas,zure2,lks,del,ks1}.
The constant coefficient $c_n$ depends on the value of $n$ and also can
 include the correction factor for including the helium in addition to
 hydrogen. $c_n$ can vary in the range of ${\cal O}(0.1\sim 10)$ for the
 parameter range of our interest \cite{mark1} and we simply set $c_n=1$
 in our analysis, which suffices for our purpose of demonstrating the
 effects of the DM-baryon coupling on the 21cm observables\footnote{We in this paper use the conventional cross section for the momentum transfer \cite{zurekyu,mark1,raby,plasma}, which is the integration of the differential cross section weighted by $(1-\cos \theta)$
\ba
\sigma (v)=\int d \cos \theta (1-\cos \theta) \frac{d \sigma(v)}{d \cos \theta}
\ea
The weight factor $(1-\cos \theta)$ is introduced to consider the
 longitudinal momentum transfer and it can regulate spurious infrared
 divergence for the forward scattering with no momentum transfer
 corresponding to $\cos \theta \rightarrow 1$.}.

We solve Eqs.~(\ref{eq:Td_z}) and (\ref{eq:Tb_z}) with
$T_\gamma = T_0 (1+z)$, where $T_0=2.73~$K, numerically. In the early
stage of the universe, it is well-known that the baryon temperature is tightly coupled with the
CMB temperature, $T_b \sim T_\gamma$. Similarly, for a sufficiently large $K_b$, the difference between $T_d$ and $T_b$ can become small in the early universe. To numerically calculate the evolution accurately
in both of these tight coupling regimes, it is useful to expand 
Eqs.~(\ref{eq:Td_z}) and (\ref{eq:Tb_z}) up to the first order in the
temperature differences as performed in Ref.~\cite{Scott:2009sz}. For this purpose, we introduce two heating time-scales due to  Compton scattering and
 DM-baryon coupling, $t_C=m_e/2\mu_b
K_{\gamma}$ and $t_{DB} = (m_d+m_{\rm H})/2 m_d K_b$, and we classify the
thermal evolution in the early universe in three cases.
The first  is the case 
with $H t_C \ll 1 $ and $Ht_{DB}\ll 1$, that is, $T_{b}$ and $T_d$ are
tightly coupled with $T_\gamma$. The second case  is for $H t_C \ll 1 $ and
$Ht_{DB} >1$, in which only $T_b$ is tightly coupled with $T_\gamma$.
The third is for $H t_C > 1 $ and
$Ht_{DB} \ll 1$ (which corresponds to $z \lesssim z_{\rm dec}$ for the parameter range of our interests as explicitly shown below).

\subsection{Regime I: $H t_C \ll 1 $ and $Ht_{DB} \ll 1$}
\label{sec:two_tight}

When $H t_C \ll 1 $ and $Ht_{DB} \ll 1$, the
difference among $T_b$, $T_d$ and $T_\gamma$ would be very small, and we can expand $T_b$ and $T_d$ as
\ba
T_b &=& T_\gamma - \epsilon_\gamma,
\label{eq:Tb_ep}
\\
T_d &=& T_b - \epsilon_b,
\label{eq:Td_ep}
\ea
where $|\epsilon_\gamma|/T_{\gamma} \ll 1$ and $|\epsilon_b|/T_b \ll 1$.
We also assume
that $\epsilon_\gamma/T_\gamma $ and $\epsilon_b/T_b$ are of the same order as
$H t_C $ and $Ht_{DB}$.

Substituting Eq.~(\ref{eq:Tb_ep}) into Eq.~(\ref{eq:Tb_z}), we obtain up to  first order in $\epsilon_b$
\begin{equation}
 \frac{\epsilon_b}{T_\gamma} =Ht_{DB},
  \label{eq:epsilon_b}
\end{equation}
where we used $T_\gamma \propto (1+z)$. Because the coefficient $1/Ht\gg
1$ is very large, we treat $d \epsilon/dz=0$ so that
$dT_d/dz=dT_b/dz=dT_{\gamma}/dz$ at  first order. Similarly
$\epsilon_\gamma$ is given by
\begin{equation}
 \frac{\epsilon_\gamma}{T_\gamma} =
  \left(1+ \frac{1}{f} \right)
  Ht_C,
  \label{eq:epsilon_gamma}
\end{equation}
where $f$ is $f=m_d \Omega_b/\mu_b
\Omega_d$.

With these approximations at hand, in terms of $\epsilon_\gamma$ and
$\epsilon_b$, the time evolutions of the temperature can be rewritten as
\ba
 \frac{d T_b}{dz}
% &=& \frac{d}{dz} (T_\gamma - \epsilon_\gamma) 
%\approx \frac{T_\gamma}{1+z} - \left(1+ \frac{1}{f} \right) \frac{T_\gamma}{1+z}Ht_{C} - \left(1+
%  \frac{1}{f} \right) T_\gamma t_C \frac{d H}{dz} - T_\gamma H
%\left(1+ \frac{1}{f} \right) \frac{dt_C }{dz}
%\nonumber
%\\
&\approx &\frac{T_\gamma}{1+z} - \epsilon_\gamma \left( \frac{1}{1+z} +
\frac{d \ln H}{dz} + \frac{d\ln t_C }{dz} 
\right),
\label{eq:Tb_tight_1}
\\
 \frac{d T_c}{dz}
% &=& \frac{d}{dz} (T_\gamma - \epsilon_\gamma-\epsilon_b) 
%\nonumber \\
 &\approx&
\frac{T_\gamma}{1+z} - \epsilon_\gamma \left( \frac{1}{1+z} +
\frac{d \ln H}{dz} + \frac{d \ln t_C }{dz}
\right)
 - \epsilon_b \left( \frac{1}{1+z} +
\frac{d \ln H}{dz} + \frac{d\ln t_{DB}}{dz} 
\right),
\label{eq:Td_tight_1}
\ea
where we assume that $f$ is constant\footnote{Since $f$ depends on the ionization rate
through $\mu_b$, this assumption is invalid during the epochs of
recombination and reionization. We, however, checked that, even though the evolution of $f$ itself is not negligible, its effects on the temperature evolution is negligible even during these epochs.}.
The evolutions of $T_b$ and $T_d$ are obtained by solving
Eqs.~(\ref{eq:Tb_tight_1}) and (\ref{eq:Td_tight_1}) with
Eqs.~(\ref{eq:epsilon_b}) and (\ref{eq:epsilon_gamma}).

\subsection{Regime II: $H t_C \ll 1 $ and $Ht_{DB} > 1$}\label{sec:caseII}

Although the DM temperature $T_d$ decouples from the baryon temperature $T_b$,
$T_b$ still couples with $T_\gamma$. We hence can assume that
\ba
T_b &=& T_\gamma - \epsilon_\gamma,
\label{eq:Tb_ep_2}
\ea
with $|\epsilon _\gamma |/T_\gamma \ll 1$.

Eq.~(\ref{eq:Tb_z}) provides to first order in $\epsilon_\gamma$
\begin{equation}
 \frac{\epsilon_\gamma}{T_\gamma} =Ht_{C } +\frac{t_C}{f t_{DB}}
  \left(1- \frac{T_c}{T_\gamma}\right).
\end{equation}
The redshift derivative of $T_b$ can then be approximated as
\begin{eqnarray}
\frac{d T_b}{d z}
% &=& \frac{d}{dz}(T_\gamma -\epsilon_\gamma)
% = \frac{T_\gamma}{(1+z)}-\frac{d}{dz}\left[
% T_\gamma Ht_C+ T_\gamma
% \frac{t_C}{ft_{DB}}
% \left(1-\frac{T_c}{T_\gamma}\right) \right]
% \\
&\approx&\frac{T_\gamma}{1+z} -\frac{T_\gamma}{1+z}Ht_{C} -T_\gamma t_C \frac{d
 H}{dz}
 - T_\gamma H \frac{dt_C }{dz}
 -\frac{d}{dz}\left[
 T_\gamma
 \frac{t_C}{ft_{DB}}
 \left(1-\frac{T_c}{T_\gamma}\right) \right].
\label{eq:T_b_caseII}
\end{eqnarray}

We numerically calculate the thermal evolution of $T_b$
and $T_d$ from Eq.~(\ref{eq:T_b_caseII}) along with
Eq.~(\ref{eq:Td_z}).

\subsection{Regime III: $H t_C > 1 $ and $Ht_{DB} \ll 1$}
\label{sec:baryon_decouple}

In this case, while the baryon temperature $T_b$ is already decoupled
from the CMB temperature $T_\gamma$, the dark matter temperature $T_d$ 
is coupled to $T_b$. 
We can write the dark matter temperature as
\ba
T_d &=& T_b - \epsilon_b,
\label{eq:Tc_ep_2}
\ea
with $|\epsilon_b|/T_b \ll 1$.
From Eqs.~(\ref{eq:Td_z}) and (\ref{eq:Tb_z}),
we obtain to  first order in $\epsilon_b$ and $Ht_{DB}$,
\begin{equation}
 \epsilon_b = -\left(1+ \frac{1}{f}\right)^{-1} \frac{t_{DB}}{t_C}\left(T_b-
	       T_\gamma \right).
  \label{eq:epsilon_b_2}
\end{equation}
Therefore, in this tight-coupling regime,
the evolution of $T_d$ can be approximated as
\begin{equation}
\frac{d T_d }{dz} \approx
\frac{d T_b}{dz}
- 
\epsilon_b
\left[ \frac{d \ln t_{DB}}{dz} -\frac{d \ln t_C}{dz}
 +\frac{1}{T_b-T_\gamma}
 \left(\frac{d T_b}{dz} - \frac{T_\gamma}{1+z}\right)
\right].
\end{equation}

On the other hand, the evolution of $T_b$ can be written as
\begin{equation}
%(1+z) \frac{d T_b}{dz} \approx {2} T_b +\left(
%\frac{f}{f+1}	\right)
% \frac{2\mu_b}{m_e} \frac{K_{\gamma} }{H}(T_b-T_{\gamma}).
(1+z) \frac{d T_b}{dz} \approx {2} T_b +\left(
1+\frac{1}{f}	\right)^{-1}
 \frac{1 }{H t_C}(T_b-T_{\gamma}).
\end{equation}
Since $f \propto m_d /m_{\rm H}$, the change of $T_b$ due to the
DM-baryon coupling becomes bigger for a bigger $m_d$ (with a fixed $\Omega_d$), and, in the limit of $m_d \gg m_{\rm
H}$, the baryons and DM can be described as
a single gas. In such a tight coupling limit with $m_d \gg m_{\rm
H}$, the total number density of the
DM-baryon mixed gas does not change from that of the baryon gas, and the
evolution of $T_b$ along with a large $m_d$ is similar
to the $T_b$ evolution without the DM-baryon coupling.
In other words, a small $m_d$~($\ll m_{\rm H}$) leads to a significant
increase of the total number density of the mixed gas, and the Compton
cooling term to couple $T_b$ to $T_{\gamma}$ effectively becomes
small. Hence, for a smaller $m_d$, the deviation of $T_b \approx T_d$
from $T_{\gamma}$ with the DM-baryon coupling becomes bigger compared with the deviation of $T_b$ from  $T_{\gamma}$ without DM-baryon coupling.

\section{Numerical results for DM and baryon temperature evolution}
\label{sec:thermal evo}

Following the previous section on numerical treatments of tight coupling regimes, we numerically calculate the DM and
baryon temperatures, $T_d$ and $T_b$, modifying the public code {\tt
RECFAST}~\cite{Seager:1999bc}. Before presenting the results with
different couplings between DM and baryons, we note that there exist
strong constraints on this DM-baryon coupling notably from the CMB and
large-scale structure due to the suppression of the matter density
perturbations where the DM perturbation growth is suppressed because of
the drag force arising from the momentum transfer between the DM and
baryon fluids \cite{chen, mark1,wand,davidson2}.
For instance, small-scale observations~(Lyman-$\alpha$ forest) by SDSS and
the CMB data by Planck can set upper bounds on the coupling between DM and baryons of order 
$\sigma_0/m_{d} \lesssim 10^{-17,-9,-6,-3,+4}~{\rm cm}^2/{\rm g}$ 
for $n=-4,-2,-1,0,+2$ \cite{mark1}. For the purpose of presenting our
findings through a concrete example, in the following we discuss the
scenarios of $n=-4$ (typical for a millicharged DM scenario \cite{mel,zurekyu,davidson,feld,zure2}) because a large negative power leads to a prominent enhancement in the cross section for a smaller momentum transfer at low redshift. We found, for the scenarios with $n=-2,-1,0,+2$, that the DM-baryon coupling cannot lead to any appreciable change in the 21 cm power spectrum within the aforementioned cross-section upper bounds from the currently available data.  

\begin{figure}
 \centering
\begin{minipage}{.49\columnwidth}
 \centering
   \includegraphics[keepaspectratio, width =0.9\linewidth]{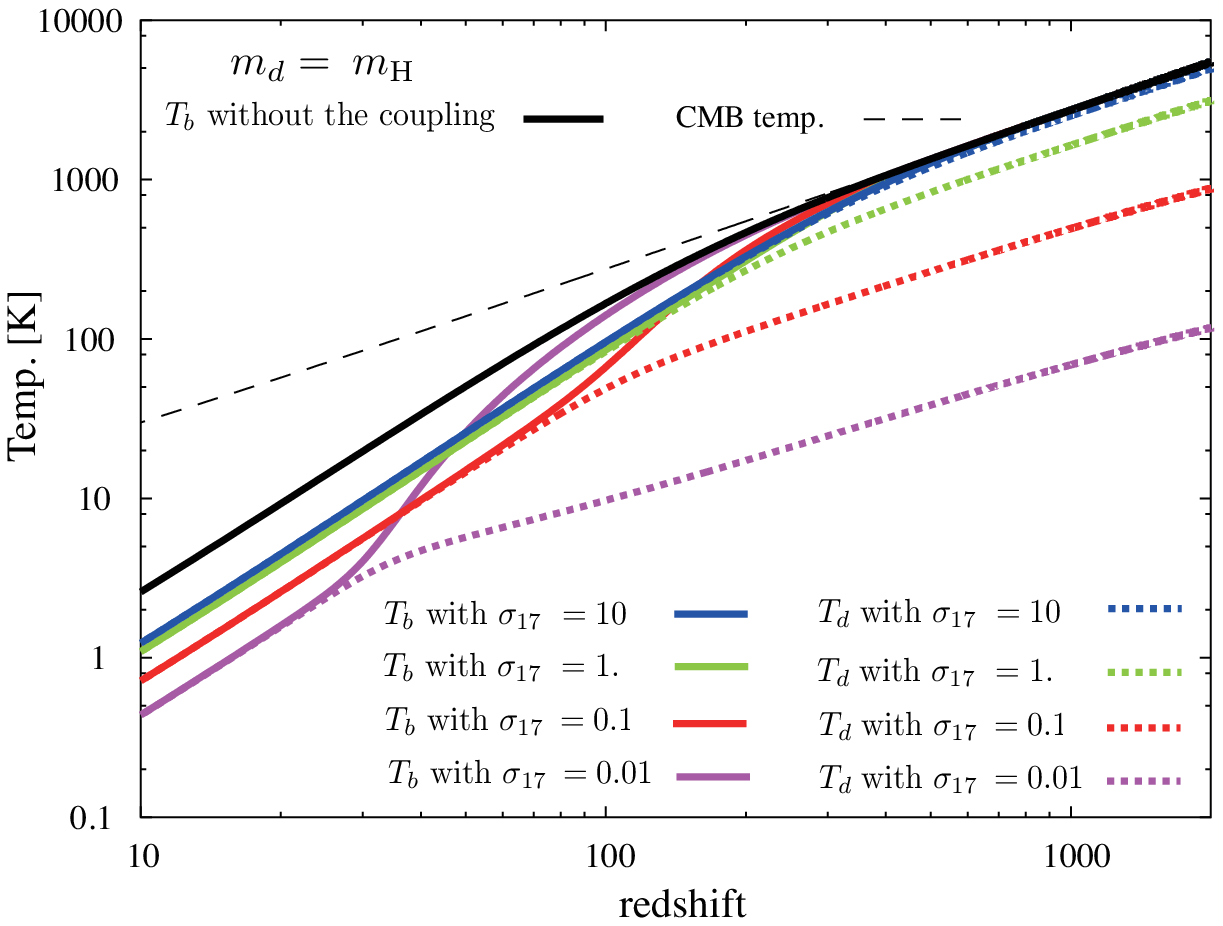}
 \end{minipage}
 \begin{minipage}{.49\columnwidth}
   \centering
   \includegraphics[keepaspectratio, width =0.9\linewidth]{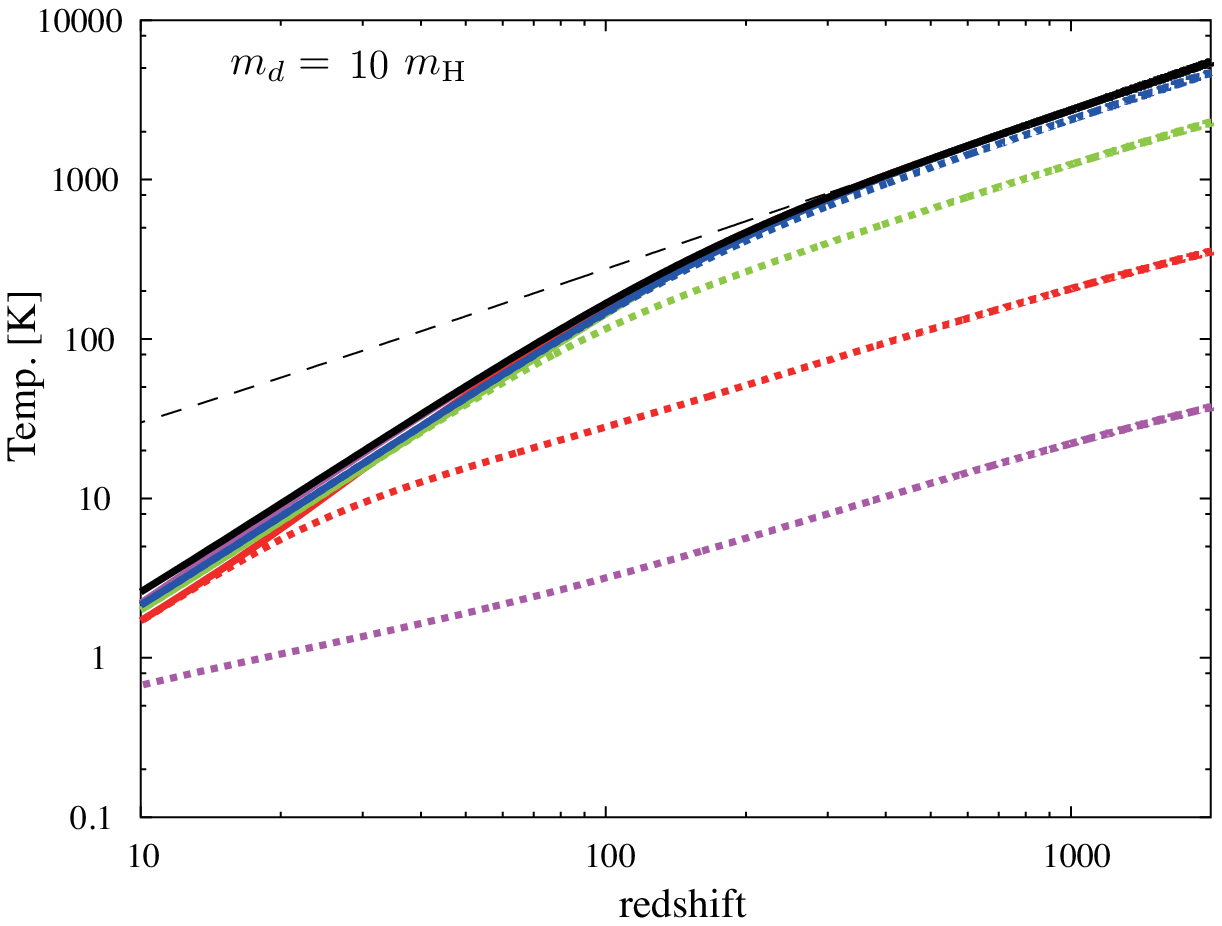}
 \end{minipage}
  \caption{
The baryon and dark matter temperature evolution for
  different values of DM-baryon coupling (the DM-baryon elastic
 scattering cross section is parameterized as $\sigma=\sigma_0 v^{-4}$,
 with $\sigma_0=\sigma_{17}m_{\rm H} 10^{-17}{\rm cm^2/g}$). We set $m_d=m_{\rm H}$ in the left panel and
 $m_d = 10 m_{\rm H}$ in the right panel. The solid and dotted lines represent the
  baryon and dark matter temperatures, respectively.
The CMB temperature
  is plotted as the dashed line.
  The magenta, red, green and blue lines are for $\sigma_{17} =0.01$,
  0.1, 1.0 and 10 respectively.
 The black solid line shows the baryon temperature evolution without DM-baryon coupling~($\sigma_{17}=0$).}
  \label{fig:thermal}
\end{figure}

Fig.~\ref{fig:thermal} represents
the temperature evolution with $n=-4$ for different values of $\sigma_{17}$, where we normalized the coupling constant as $\sigma_{0} = \sigma_{17}
m_{\rm H}
\times  10^{-17}~{\rm cm^2/g}$. 
To demonstrate the mass dependence, we simply show the results for $m_d = m_{\rm H}$ and $10 m_{\rm H}$ in Fig.~\ref{fig:thermal} for different values of DM-baryon coupling. %% The upper bound in the limit of $m_d \gg m_{\rm H}$, $\sigma_0/m_d <
%% 10^{-17} ~{\rm cm^2/g}$, obtained in Ref.~\cite{mark1} corresponds to the
%% line for $\sigma_{17} = 10$ in the right panel.
%With the values $\sigma_{17}$ in Fig.~\ref{fig:thermal},
At high redshifts, $z > z_{\rm dec}$, $T_b$ is tightly coupled to $T_{\gamma}$ ($T_{b}\approx T_{\gamma}$, and hence the thermal evolution can be described with the treatment in in Sec.~\ref{sec:caseII} where $T_d\propto 1/t_{DB}$). It is consequently difficult to find any difference between
the evolution of $T_b$ for the different couplings in
Fig.~\ref{fig:thermal} at  high redshift. Note, however, that $T_d$
deviates from $T_b \approx T_{\gamma}$ at high redshifts. In the presence
of DM-baryon coupling, the DM thermal evolution is not adiabatic and is determined by the balance between
the adiabatic cooling and the heating due to the coupling. We can infer,
by substituting $T_b  \approx T_\gamma$ in Eq.~(\ref{eq:Td_z}), that DM
evolution follows $T_d \sim T_\gamma /t_{DB} H$. More precisely, from
Fig.~\ref{fig:thermal}, we numerically find that the DM temperature is
well approximated by the fitting formula $T_d \approx T_\gamma /1.5
t_{DB} H$. The time-scale $t_{DB}$ is proportional to $(m_d+m_{\rm
H})^2/\sigma_{17}m_d$. When $m_d \gg m_{\rm H}$, $t_{DB}\propto m_d$ which results in $T_d\propto 1/t_{DB}\propto 1/m_d$, and Fig.~\ref{fig:thermal} indeed shows that $T_d$ is larger for a smaller $m_d$. 

Let us here note that the DM-baryon momentum transfer rate $K_b$ given
in Eq. \ref{eq:thermal_coupling}, hence the thermal evolution at high
redshifts,  turns out to be heavily dependent on $T_b$ but not so much
on $T_d$, where the  temperature dependence of $K_b$ shows up in the
factor $(T_b/ m_{\rm H}+T_d/m_d)$. For $m_d \gg  m_{\rm H}$, $(T_b/ m_{\rm H}+T_d/m_d)\sim
T_b/ m_{\rm H}$ to leading order in $ m_{\rm H}/m_d$. For $m_d \ll
m_{\rm H}$ on the other hand, $T_d \sim 1/t_{DB} \sim m_d$ and the $m_d$
dependence cancels out in $T_d/m_d$\footnote{Consequently, because the
baryon temperature never exceeds the cold dark matter temperature, this
factor $(T_b/ m_{\rm H}+T_d/m_d)$ would be at most of order $\sim 2
\times T_b/ m_{\rm H}$ saturated at $m_d \sim  m_{\rm H}$. We hence
expect the upper bound $\sigma_0 \lesssim 10^{-16} m_{\rm H}~{\rm cm^2/g}$ (corresponding to $\sigma_{17}=10$ in our notation) which Ref. \cite{mark1} obtained for $m_d= 10$ GeV would not become significantly tighter even for a smaller dark matter mass. We therefore restricted the parameter range of our discussion to be $\sigma_{17}\leq 10$ and presented the results for $m_d=10,1$ GeV, which would suffice our purpose of showing the potential significance of the DM-baryon coupling on the 21 cm signals.}.

At  low redshifts after $T_b$ has decoupled from $T_{\gamma}$, $z \lesssim z_{\rm dec}$, the coupling between baryons and dark matter affects the temperature
evolution of baryons. 
The baryons become cooler through the DM-baryon coupling because $T_d<T_b$ as compared with no-coupling scenarios. Sufficient coupling can make the
temperatures of baryons and DM equal. Once they match each other, the coupling term in the Boltzmann equations ($\propto (T_b-T_d)$) reaches effectively zero and the thermal evolution becomes adiabatic, that is, $T_b$ and $T_d$ are
proportional to $(1+z)^{-2}$ because the DM and baryons have the same
adiabatic index. Since we set $n=-4$ for the velocity-dependence of the coupling,
the coupling strength becomes bigger for a smaller momentum transfer at a smaller redshift. The evolution of $T_b$ is modified at lower redshifts
even for a small $\sigma_{17}$ for $m_d=m_H$ in the left panel. We find, however, that, when
$\sigma_{17}< 0.001$, the baryon temperature does not couple with
the dark matter temperature even at lower redshifts and its evolution is similar to the case
without the coupling.
The DM-baryon coupling term for the baryon temperature evolution, which appears in
Eq.~(\ref{eq:Tb_z}), becomes small with increase of $m_d$, as confirmed in Fig.~\ref{fig:thermal}.
% Consequently, compared to the
%case for $m_d = m_{\rm H}$, the baryon temperature is relatively less  coupled to the
%dark matter temperature for a bigger $m_d=10m_H$.

For a sufficiently large value of DM-baryon coupling (in our example, for $\sigma_{17}>10$),
the DM temperature is well coupled with the baryon temperature, and $T_d
\approx T_b$ is established even around the epoch when the baryon
temperature starts to decouple from the CMB temperature. The evolution
in this regime corresponds to the tight-coupling case discussed in Sec.~\ref{sec:baryon_decouple} where a small DM mass, due to a small Compton coupling between $T_b$ and $T_{\gamma}$, leads to the early decoupling of $T_b \approx T_d$ from $T_{\gamma}$. The difference of the baryon temperature evolution from  the no DM-baryon coupling scenarios hence becomes bigger for a smaller DM mass.

Finally it is worth mentioning the case in the limit of $m_d \ll m_{\rm H}$.
At high redshifts $z>z_{\rm dec}$, the time scale $t_{DB}$ is proportional to $1/m_d$, and the DM temperature $T_d \propto 1/t_{DB}$ which decreases as $m_d$ becomes small.

The DM-baryon coupling term in
Eq.~(\ref{eq:Tb_z}) does not become small in the limit of $m_d \ll m_{\rm H}$, in contrast to $m_d \gg m_{\rm H}$ case, and, in fact, becomes
independent of $m_d$ with only its dependence on $\sigma_0$. 
Hence the baryon temperature can be dragged to the lower dark matter
temperature, and one finds that the change in the $T_b$ evolution is bigger for a smaller $m_d$.

\section{The evolution of 21~cm signals with DM-baryon coupling}

The DM-baryon coupling can affect the evolution of the baryon
temperature as shown in the previous section, and the measurement of baryon temperature in the dark ages, in
particular during $20<z< z_{\rm dec}$, could well reveal the nature of DM. 
The measurement of redshifted 21~cm lines from neutral hydrogen is
expected to be a good probe of baryon gas in the dark ages. The strength
of the emission or absorption of the 21~cm lines depends on the
density, temperature and ionization fraction of baryon gas.

The observational signals of redshifted 21~cm lines are measured as the
difference between the brightness temperature of redshifted 21~cm signals and
the CMB temperature. 
This differential brightness temperature is given by
\begin{equation}
\delta T_b (z) = \left[
  1-\exp (-\tau)\right] \frac{T_s -T_\gamma}{1+z},
\label{eq:def_dTb}
\end{equation}
where $\tau$ is the optical depth and $T_s $ is the spin temperature.
The spin temperature describes the number density ratio of hydrogen atoms
in the excitation state to those in the ground state, and is given by~\cite{Field1958,Field1959}
\begin{equation}
 T_s = \frac{T_*+T_\gamma +y_k T_b}{1+y_k},\label{eq:spin}
\end{equation}
%where $y_k$ represents the kinetic coupling term and, since we are
where $T_*$ is the temperature corresponding to the energy of hyperfine
structure of neutral hydrogen and $y_k$ represents the kinetic coupling term given by
\begin{equation}
 y_k = \frac{T_*}{AT_b}(C_{\rm H}+C_e +C_p),
\end{equation}
where $A$ is the spontaneous emission rate and $C_{\rm H} $, $C_e  $, and $C_p$ are
the de-excitation rates of the triplet due to collisions with neutral atoms, electrons, and protons \cite{fur}.
For these rates, following Ref.~\cite{Lewis:2007kz}, we adopt the values from
Refs.~\cite{fur,Furlanetto:2007te}.
Since we are
interested in the signals from the dark age, we neglect the
Lyman-$\alpha$ coupling~(Wouthysen field effect) term~\cite{Field1958, Wouthuysen1952} in Eq.~(\ref{eq:spin}), which is
ineffective without luminous objects.

\begin{figure}
 \centering
\begin{minipage}{.49\columnwidth}
 \centering
   \includegraphics[keepaspectratio, width =0.9\linewidth]{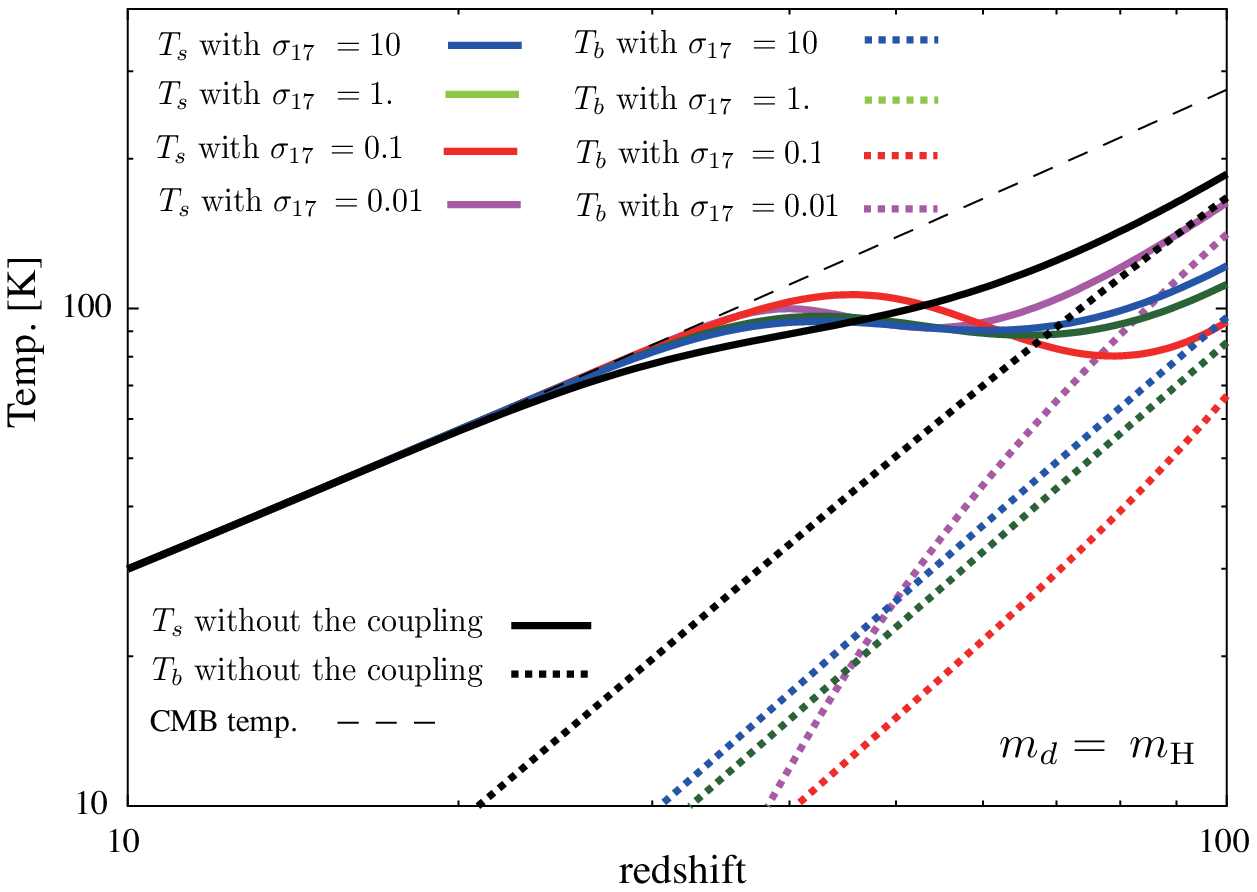}
 \end{minipage}
 \begin{minipage}{.49\columnwidth}
   \centering
   \includegraphics[keepaspectratio, width =0.9\linewidth]{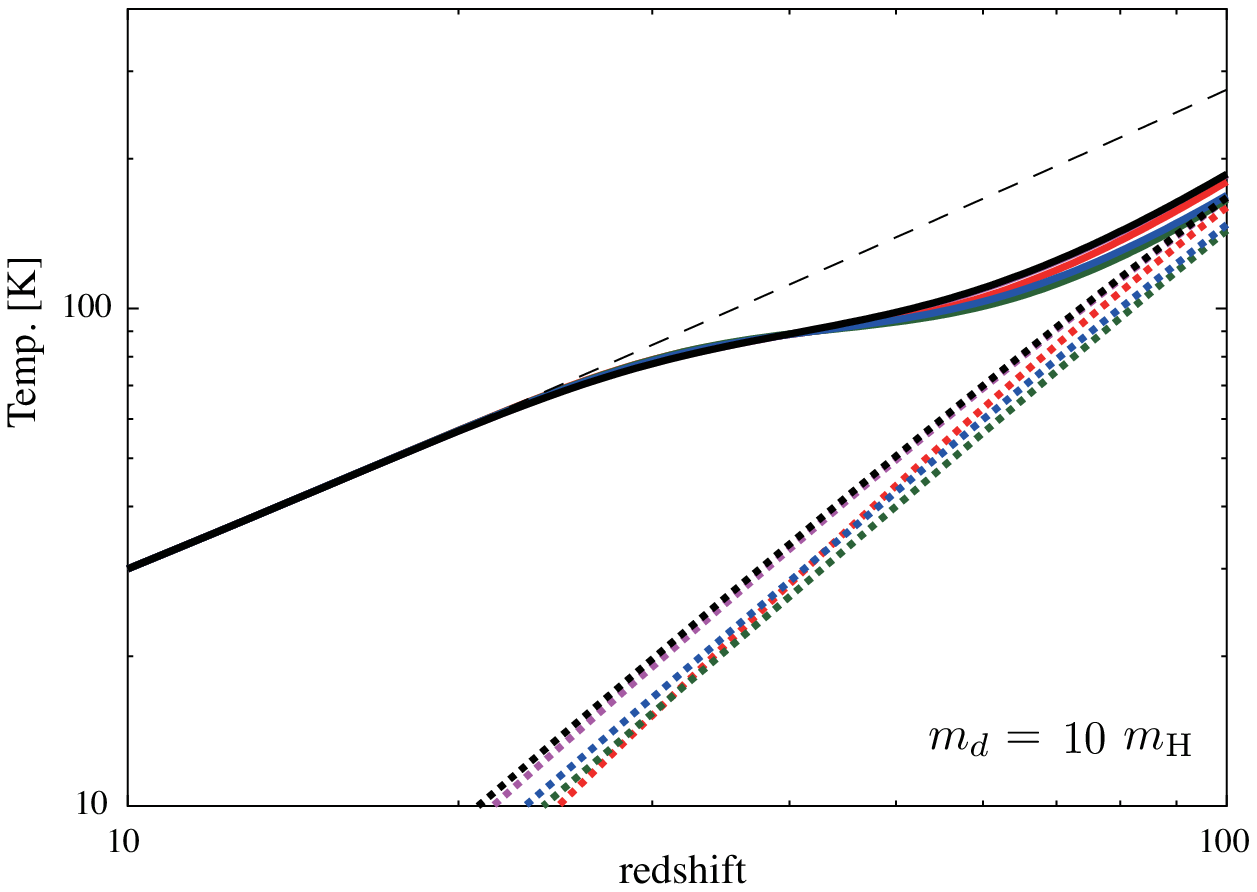}
 \end{minipage}
  \caption{The spin temperature evolution for
  different values of DM-baryon coupling. The solid, dotted and dashed lines represent the
  spin, baryon and CMB temperatures, respectively.
 We set $m_d=m_{\rm H}$ in the left panel and
 $m_d = 10 m_{\rm H}$ in the right panel. 
  The magenta, red, green and blue lines are for $\sigma_{17} =0.01$,
  0.1, 1.0 and 10 respectively.
 The temperatures evolution without DM-baryon coupling~($\sigma_{17}=0$) is
 plotted in black.}
  \label{fig:spin}
\end{figure}

We show the evolution of $T_s$ for different DM-baryon coupling values in Fig.~\ref{fig:spin}.
%The evolutions differ for different values of $\sigma_{17}$. 
As one can expect
from Fig.~\ref{fig:thermal}, the difference from the case without the
coupling is larger for $m_d = m_{\rm H}$ than for $m_d = 10 m_{\rm H}$.
The 21~cm signals depend on $T_s$, and we hence can expect the redshift evolution of the differential brightness
temperature also depends on $\sigma_{17}$.

Measurements of cosmological $21$~cm signals will be performed by
interferometers such as LOFAR and SKA which can measure the fluctuations in the
differential brightness temperature. The angular power spectrum of
$\delta T_b$ is given by
\begin{equation}
C_\ell(z) = \delta T_{b0}^2 \int dk~k^2 \Delta^2 _{21 , \ell}(z,k) P(k),  
\end{equation}
where $\Delta_{21,\ell}$ is the transfer function for the 21~cm
fluctuations, $P(k)$ is the power spectrum of the primordial
curvature perturbations and $\delta T_{b0}$ is the value of the
differential 21cm brightness temperature
which can be approximated by~\cite{Ciardi:2003hg}
\begin{equation}
 \delta T_{b0} \approx 26~{\rm mK}~ x_{H} \left(
1-\frac{T_\gamma}{T_{s}} \right) \left(\frac{h^2 \Omega_b}{0.02}\right)
\left[\left(\frac{1+z}{10}\right)\left(\frac{0.3}{\Omega_m}\right)\right].\label{eq:spin0}
\end{equation}

In this paper, since we consider the effect of the coupling between baryons
and dark matter on the temperature evolution, we focus only on the
modification of $\delta T_{b0}$ due to the coupling.
We ignore the effect of the DM-baryon coupling on the evolution of the
density fluctuations \cite{mark1, chen}.
Therefore, the transfer function
$\Delta_{21, \ell} $, which we calculate by using \texttt{CAMB }~\cite{Lewis:2007kz},
is the same as that in the standard $\Lambda$CDM model.

\begin{figure}
\centering
\begin{minipage}{.49\columnwidth}
	\centering
	\includegraphics[keepaspectratio, width=0.8\linewidth]{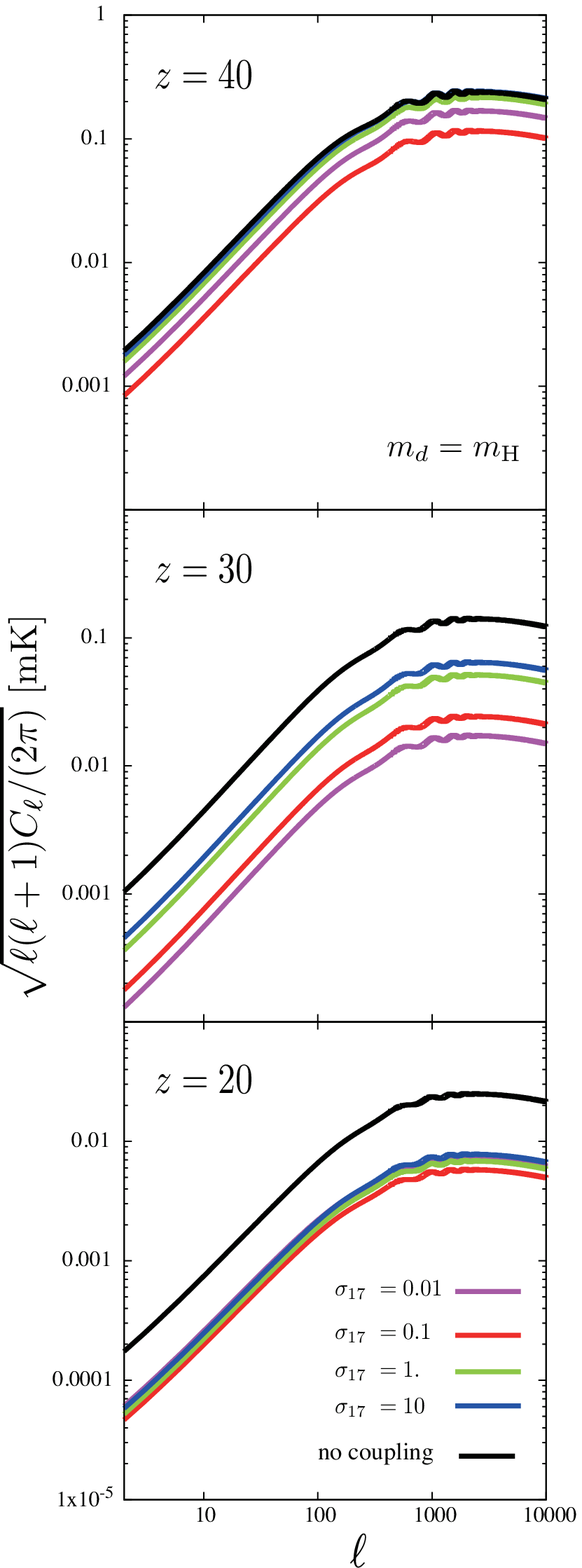}
\end{minipage}
\begin{minipage}{.49\columnwidth}
 	\centering
	\includegraphics[keepaspectratio, width=0.8\linewidth]{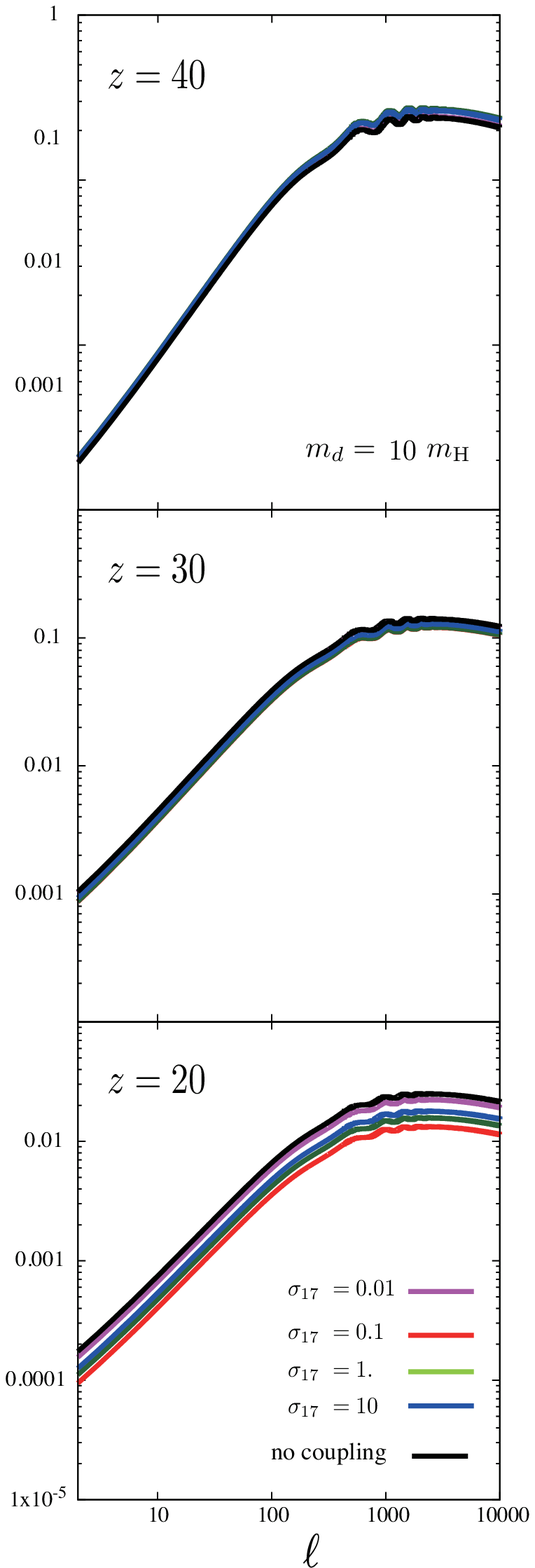}
\end{minipage}
	\caption{The evolution of the angular power spectrum $C_\ell$
 for different values of DM-baryon coupling. We set $m_d =m_{\rm H} $ and $10m_{\rm
 H}$ in the left and right panels, respectively.
In both left and right panels, 
 the redshifts are set to
 $z=40$, 30 and 20 from the top to bottom, respectively.}
\label{fig:clplot}
\end{figure}

We show the
dependence of the angular power spectrum $C_\ell(z)$ on DM-baryon coupling in Fig.~\ref{fig:clplot}.
According to Eq.~(\ref{eq:spin0}), the evolution of $\delta T_{b0}$ depends on
the spin temperature shown in Fig.~\ref{fig:spin}.
The coupling between baryons and
dark matter lowers the baryon temperature. Therefore, the kinetic
coupling term for the hyper-fine structure in Eq.~(\ref{eq:spin})
becomes small due to the low baryon temperature. The spin
temperature then quickly approaches the CMB temperature for $z \lesssim 50$, which results in a smaller amplitude of $C_\ell$ compared with the no coupling case.
For instance, for $\sigma_{17} < 0.1$ with $m_d = m_{\rm H}$, the amplitude of
$C_\ell$ is suppressed by $1/10$~(see the red and magenta lines in the left panel of
Fig.~\ref{fig:clplot}). As the coupling increases, the dark matter
temperature becomes larger and approaches the  baryon temperature as shown in Fig.\ref{fig:thermal}.
 Fig.~\ref{fig:clplot} indeed shows that the amplitudes are comparable, except in the small coupling case ($\sigma_{17}=0.01$). The behavior for this small cross-section is due to the fact that $T_b$ turns out not to couple with $T_d$ at high redshifts $z > 50$ due to the small coupling. As $T_b$ becomes smaller at lower redshifts, however, the coupling can become more effective due to the enhancement for small momentum transfer.
%% The behavior for a small cross section $\sigma_{17}=0.01$ 
Fig.~\ref{fig:clplot} confirms our expectation that the effects of DM-baryon coupling on the $T_b$ evolution becomes small as $m_d$ increases (as mentioned at the end of \S \ref{sec:baryon_decouple}).
%% Therefore, when $m_d = 10 m_{\rm H}$, the suppression of $C_\ell$ due to the coupling is almost negligible
%% at relatively high redshifts $z=40$ and $30~$(see the right panels of Fig.~\ref{fig:clplot}).
%% At a low redhisft $z=20$, although one can see the suppression of $C_\ell$ depending on
%% $\sigma_{17}$, the suppression is still small compared with the cases for
%% a smaller $m_d = m_{\rm H}$

Note that, while the amplitude of $C_\ell$ is suppressed due to the
coupling between baryons and dark matter at a low redshift ($z \lesssim 40$), it is amplified at a high redshift ($z \gtrsim 50$). 
This is because, at high redshifts, the kinematic coupling term in Eq.~(\ref{eq:spin}) is significant and the spin
temperature is tightly coupled with the baryon temperature. The deviation of the spin temperature from the CMB
temperature hence becomes large and $C_\ell$ is consequently amplified at high redshifts.

Let us also comment on $C_\ell$ when $m_d$ is smaller than $m_{\rm H}$.
As mentioned in Sec.~\ref{sec:thermal evo},
the baryon temperature is strongly dragged to the dark matter
temperature which becomes small with decreasing  $m_d$.
Therefore, the kinetic coupling term is small for a small $m_d$ and
the spin temperature has a tighter coupling with the CMB temperature. 
This tight coupling causes the strong suppression of $C_\ell$, according
to Eq.~(\ref{eq:spin0}).
As a result,
when $m_d \ll m_{\rm H}$, the suppression due to the coupling is
significant even at large redshifts.

\section{Discussion and conclusion}
Before concluding our studies on 21 cm signals, let us briefly mention other relevant observables which could potentially be affected by the change in the background temperature evolution because of the DM-baryon coupling.
%\subsection{CMB anisotropies}

\begin{figure}
 \begin{center}
  \includegraphics[keepaspectratio, scale =0.7]{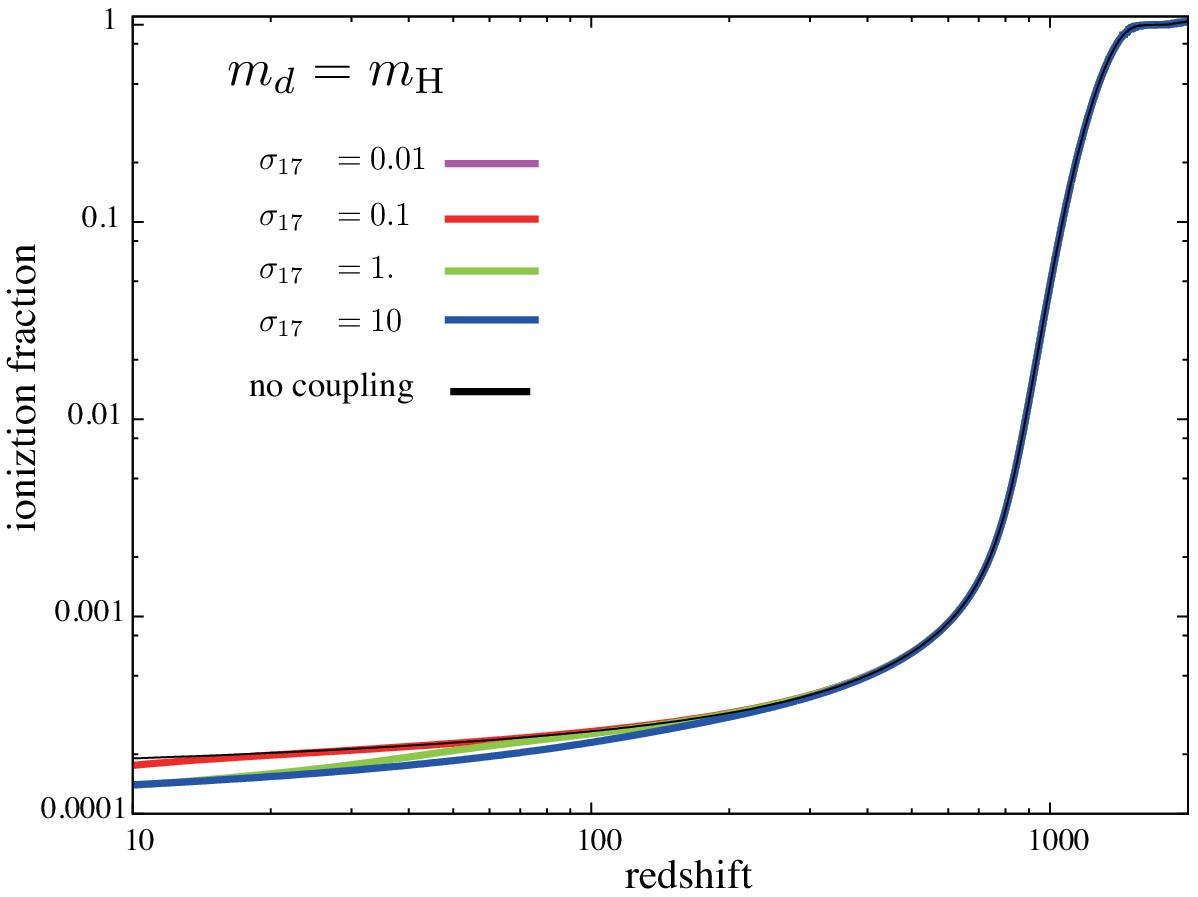}
  \caption{The evolution of the ionization fraction for different values of DM-baryon coupling.}
  \label{fig:ion}
 \end{center}
\end{figure}

{\bf Epoch of recombination and CMB anisotropies:~}If the baryon temperature changes around the epoch of recombination, 
the last scattering surface could be modified and this modification can produce a
footprint on the CMB temperature anisotropies.
We evaluate the ionization fraction for different
$\sigma_{17}$ and plot the results in Fig.~\ref{fig:ion}.
We found, since the baryon temperature is strongly coupled with the CMB
temperature around these redshifts~(see Fig.~\ref{fig:thermal}), the dark matter cooling cannot
decrease the baryon temperature enough to modify the epoch of
recombination.
Therefore, the coupling between baryons and dark matter cannot produce a
observable signature in the primordial CMB anisotropies.

At lower redshifts, when the baryon temperature decouples from the
CMB temperature, the dark matter cooling could affect the
thermal evolution of baryons. Since the baryon temperature is dragged to
lower temperature, the residual ionization fraction becomes
small. It is, however, difficult to measure such small residual
ionization fraction by cosmological observations.

%\subsection{CMB distortions}
{\bf CMB distortions:~}
Precise measurements of CMB spectral distortions from the blackbody
spectrum can be a promising probe of thermal history of the Universe~(see Ref.~\cite{Tashiro:2014pga} for a recent
review).
Generally CMB distortions can be classified into two types~\cite{zeldovich1969,sunyaev1970}. One is the $\mu$-type
distortion which is generated between $z\sim 10^6$ and $z \sim
10^4$. The other is the $y$-type distortion which is produced after the
epoch of the $\mu$-type distortion generation~($z \lesssim 10^4$).

The difference in the adiabatic indexes between baryons and
CMB photons can create CMB distortions~\cite{Khatri:2011aj}.
Because the baryon temperature is always lower than the CMB temperature,
the energy of the CMB photons is transferred to baryons via Compton scatterings.
This energy transfer modifies the CMB frequency spectrum and we can observe
this modification as CMB distortions.
Following Ref.~\cite{Khatri:2011aj},
in order to evaluate the CMB
distortions due to this baryon cooling, it is useful to define the
parameter $Y_{\rm BEC}$ as
\begin{equation}
 Y_{\rm BEC} = -\int dz \left(1- \frac{T_b}{T_{\gamma}}\right)
  \frac{k_B \sigma_T}{m_e c} \frac{n_e T_\gamma}{(1+z) H}.
\end{equation}
For example, the $y$-parameter which characterizes the $y$-type distortion is obtained
by $y = - Y_{\rm BEC}$.

We evaluated $Y_{\rm BEC}$ for different values of $\sigma_{17}$. We find that
$Y_{\rm BEC}$ becomes at most ${\cal O}(10^{-9})$ for the parameter range of  interest, $10^{-3} < \sigma_{17}< 10^2$, while $Y_{\rm BEC}$ is on
the order of $10^{-10}$ without the coupling between baryons and dark
matter.
The value of  $Y_{\rm BEC}$ corresponds to $\mu \sim 10^{-9}$ for the
$\mu$-type distortion and $y \sim 10^{-9}$ for the $y$-type distortion.
Because the Silk damping of the primordial density perturbations
produces $\mu \sim 10^{-8}$~\cite{Dent:2012ne, Chluba:2012we} and the reionization process gives $y \sim
10^{-7}$~\cite{Sunyaev:2013aoa}, it would be difficult to find the signature of the coupling between
baryons and dark matter in the CMB distortions.
\\
\\
%\section{discussion and conclusion}
We have demonstrated that DM-baryon coupling can affect the background temperature evolution and consequently the 21~cm signal. Our specific example, the velocity-dependent elastic scattering cross-section, would be also of great interest for particle physics studies because of its infrared enhancement for a low momentum transfer, which has been explored for potential signals beyond the standard model at  collider and dark matter search experiments \cite{feld,zure2,lks,del,ks1}. Such probes of the dark matter properties from both cosmology and particle physics deserve further study in view of forthcoming experiments which can explore the nature of the DM coupling to  ordinary baryons.

We have shown that the 21~cm signal is suppressed due to
the existence of  DM-baryon coupling, and it would certainly be useful to provide  further constraints on DM-baryon coupling. For instance, we have found
that the 21 cm brightness temperature angular power spectrum can be
suppressed by a factor 2 for $m_d=10$ GeV within the current bounds from
the CMB and Ly$-\alpha$ data. This overall suppression can even be larger
for a smaller dark mass with a fixed  cross-section, for instance of
order a factor 10 for $m_d=1$ GeV. We have however found  that the degree of
further suppression becomes milder for an even smaller $m_d \ll m_{\rm H}$, partly because the temperature dependence of the DM-baryon momentum
transfer rate $K_b$ on the dark matter mass  saturates at $m_d\sim
m_{\rm H}$ and becomes independent of $m_d$ for $m_d \ll m_{\rm H}$. 
%suppression reaches about one tenth for $m_d <1~{\rm GeV}$ even with the current upper bound.

We plan to explore the effects of DM-baryon coupling on the evolution of fluctuations in  future work where one needs  extra care in the treatment of non-linearities. Some simplifications made in our analysis would also deserve further study. For instance, we considered only the thermal velocity and did not include the peculiar velocity contributions in estimating the DM-baryon momentum transfer rate. 
Even though the inclusion of such bulk velocity contributions does not always change the constraints on the upper bounds on the allowed DM-baryon scattering cross sections, there are cases where the cross-section constraints could get tighter (possibly even by a factor 10) even though more detailed numerical analysis is needed because of the uncertainties caused by non-linear evolution \cite{mark1}. 
%Even though the inclusion of such bulk velocity contribution does not always change the constraints on the upper bounds on the allowed DM-baryon scattering cross section for $n=-2,0,2$, there are cases such as that for $n=-4$ where the cross section constraints could get tighter by a factor 10 even though the detailed numerical analysis is awaited due to the uncertainties caused by non-linear evolutions \cite{mark1}. 
\\
\\
\\
This work was supported by the MEXT of Japan, Program for Leading Graduate Schools "PhD
 professional: Gateway to Success in Frontier Asia'',
the Japan Society for Promotion of Science (JSPS) Grant-in-Aid for Scientiffic Research
(No.~25287057),
 the ERC project 267117 (DARK) hosted by Universit\'e Pierre et Marie Curie - Paris 6 and at JHU by NSF grant OIA-1124403.
%%%%%%%%%%%%%%%%%%%%%%%%%%%%%%

\end{document}